\documentclass[a4paper]{PoS}
\usepackage{epstopdf}
\usepackage{float}
\usepackage{subfigure}

\usepackage{xcolor}

\def\watermark#1{%
\setlength{\unitlength}{1pt} 
\begin{picture}(0,0) 
\put(120,-80){\textcolor{gray!50!white}{\Huge #1}}    
\end{picture} }

\title{Using UV-pass filters for bright Moon observations with MAGIC}

\ShortTitle{Using UV-pass filters for bright Moon observations with MAGIC}

\author{\speaker{D. Guberman}$^{a}$, J. Cortina$^{a}$, R. Garc\'ia$^{a}$, J.Herrera$^{b}$, M. Manganaro$^{b}$, A. Moralejo$^{a}$, J. Rico$^{a}$ and M. Will$^{b}$ for the MAGIC collaboration\\
        \llap{$^{a}$}Institut de F\'isica d'Altes Energies (IFAE), E-08193 Bellaterra, Spain\\
        \llap{$^{b}$}Instituto Astrof\'isico de Canarias (IAC), E-38205 La Laguna, Spain\\
        E-mail: \email{dguberman@ifae.es}, \email{cortina@ifae.es}
        }

\abstract{MAGIC is a system of two Imaging Atmospheric Cherenkov Telescopes (IACT) that observe Very High Energy (VHE) gamma ray sources. The PMTs in their cameras are designed to operate under moonlight, but they are limited to Moon phases below $93\%$ (300 Moon hours per year), as they can get damaged if the amount of light they receive is too high. As a result, they cannot be used in the three to five nights around full Moon. We have selected commercial inexpensive UV-pass filters rejecting light above a wavelength of 420 nm, where the moonlight intensity is stronger. We mounted them on light-weight frames that can be easily installed on the telescope cameras. Test observations have been performed during the last nine months, from which a moonlight transmission of about $20\%$ and a Cherenkov light transmission of about $45\%$ are estimated. This allows the observation of sources down to an angular distance of 5 degrees to the Moon during Full Moon: essentially in the whole sky and all possible moonlight conditions. Therefore, the duty cycle of MAGIC can be extended by about $30\%$, including nights when VHE observations with IACTs are currently not feasible.  Here we evaluate the preliminary performance, in terms of sensitivity and energy threshold, of the MAGIC telescopes equipped with the UV-pass filters under different moonlight intensities, as inferred from Crab Nebula observations and Monte Carlo simulations.}

\FullConference{The 34th International Cosmic Ray Conference,\\
		30 July- 6 August, 2015\\
		The Hague, The Netherlands}

\begin{document}

\section{Introduction}

MAGIC (Major Atmospheric Gamma-ray Imaging Cherenkov telescopes) consists of two 17~m diameter Imaging Atmospheric Cherenkov Telescopes (IACT). Located on the Roque de los Muchachos Observatory on the Canary Island of La Palma, Spain
($28^\circ$ N, $18^\circ$ W), at a height of 2200 m a.s.l, they are designed to observe very high energy (VHE, $\gtrsim$ 30 GeV) $\gamma$-rays\cite{MUP}.

The telescopes achieve their best performance in the absence of moonlight. In such conditions and for zenith angles below $30^\circ$, MAGIC achieves an energy threshold of $\sim 70$ GeV and an integral sensitivity above 290 GeV of $0.67 \pm 0.04 \%$ of the Crab Nebula flux\cite{MUP}. However, as the Moon is present in half of the total night-time available during a year, observations are also performed under low and moderate moonlight. About 300 hours per year are taken in such conditions. But if the background light is too high, the Photo Multiplier Tubes (PMTs) in the MAGIC cameras suffer significant ageing. Observations in slightly brighter conditions can be achieved by reducing the High Voltage (HV) in the PMTs, but their performance degrades when the gain is reduced by more than $50\%$.

We define the \textit{Dark NSB} (where NSB stands for Night Sky Background) as the brightness of the sky in the darkest situation in which observations are possible: an extragalactic region of the sky with no moonlight and at low zenith angles. By reducing the HV, observations can be performed up to a sky brightness that is 15 times the Dark NSB. As a result the telescopes are currently not operated in the three to five nights around full Moon, when no dark time is available, and they are hardly used in the three to four nights before and after full Moon.

Between July 2013 and March 2105 we tested the use of UV-pass filters to observe under extremely bright conditions (i.e. during the full Moon period). In section \ref{sec:filters} we describe the characteristics of the selected filters: their transmission curve and the mechanical setup for mounting them in front of the MAGIC cameras. In section \ref{sec:samples} we show the data samples of the Crab Nebula that were taken with UV-pass filters, that were used to estimate the performance of the telescopes with the filters. Crab Nebula is considered as the brightest steady VHE $\gamma$-ray source, normally labelled as the ``standard candle'' of the VHE $\gamma$-ray astronomy and it is frequently used to evaluate performance of VHE instruments\cite{MUP}. In section \ref{sec:performance} we evaluate the preliminary performance of the system under different brightness conditions, based on the results from the analysis of the Crab Nebula samples and Monte Carlo simulations that were specially tuned to include the effect of the filters and the higher background light. Final remarks and the implications of the evaluated performance are discussed in section \ref{sec:conclusions}.

\section{Filters design}\label{sec:filters}

\begin{figure}[t]
\centering
\includegraphics[width=.6\textwidth]{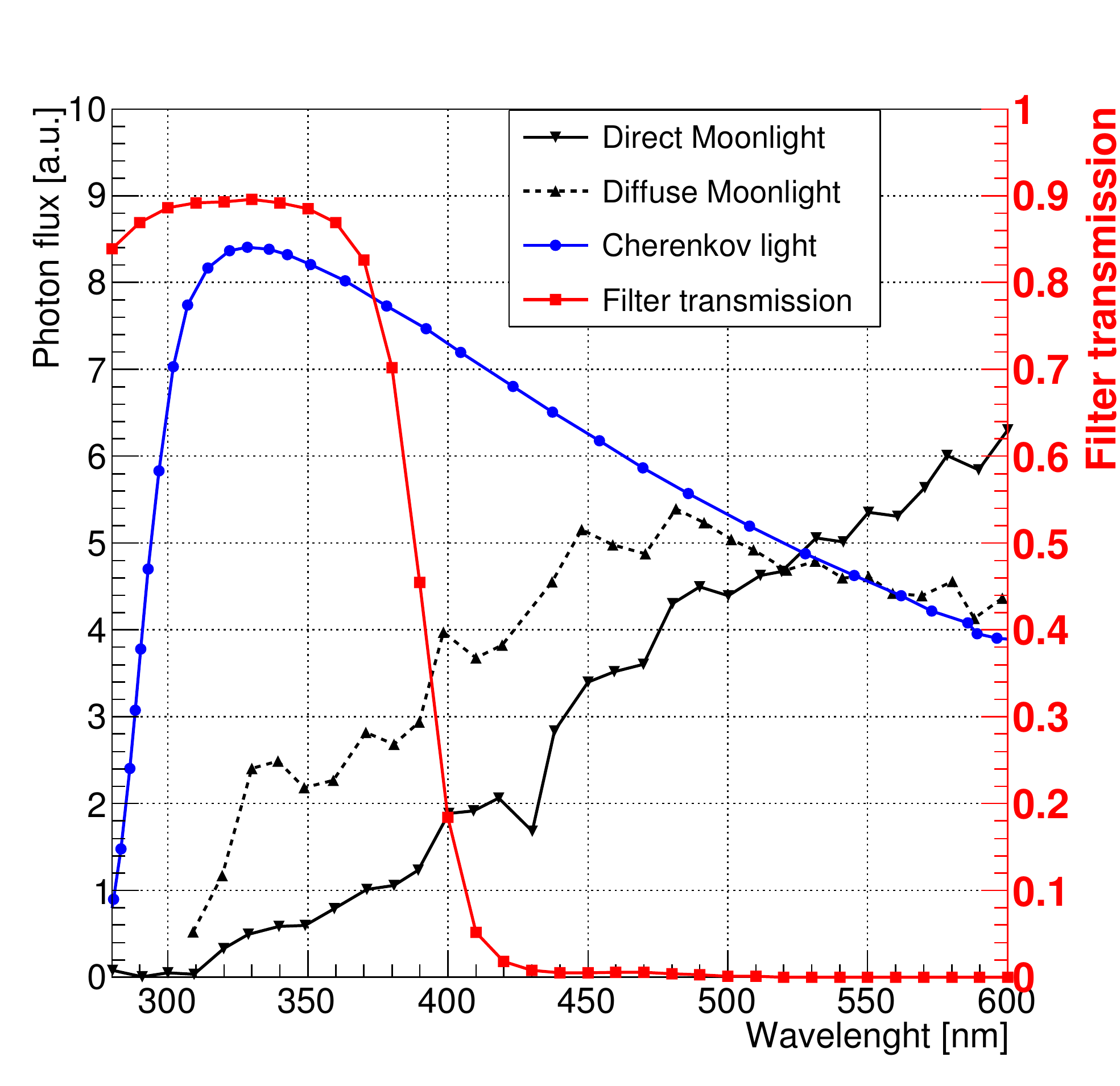}
\caption{The blue curve shows the typical Cherenkov light spectrum for a vertical shower initiated by a TeV $\gamma$-ray, detected at 2200 m a.s.l.\cite{Doering}. The black solid curve shows the shape of direct moonlight spectrum. The dashed black curve, the spectrum from diffuse moonlight. The three curves are scaled by an arbitrary normalization factor. The filters transmission curve is plotted in red.}\label{fig:transmission}
\end{figure}

We have selected filters with a transmission curve that maximizes Cherenkov light transmission and rejects as much background light as possible. In addition, the wavelength of MAGIC calibration laser is 355 nm, which already puts a constrain on where the wavelength cut can be made. During moonlight nights the spectrum of the background light that reaches the telescopes depends on the distance between their pointing position and the Moon. If they are pointing far from the Moon (tens of degrees away) the background light does not come directly from the Moon, but of Rayleigh-scattered moonlight (``diffuse moonlight'') that peaks at $\sim 470$ nm. When pointing close to the moon, Mie-scattering of moonlight dominates and its intensity is higher at higher wavelengths (``direct moonlight''). The actual shape of the spectrum depends on the aerosol content and distribution and the zenith angle of the Moon. The diffuse and direct moonlight spectra can both be obtained by folding the solar spectrum with the Albedo of the Moon. This was done using the code SMARTS\cite{SMARTS1, SMARTS2} and is shown in figure \ref{fig:transmission}.

The spectrum of Cherenkov light of showers depends mainly on the altitude of the shower maximum, but also on the nature of the incident particle (whether is a $\gamma$-ray or a hadron) and its energy. For a vertical shower initiated by a TeV $\gamma$-ray, detected at 2200 m a.s.l, it peaks at $\sim 330$ nm, as shown in figure \ref{fig:transmission}. Taking all into account, we selected commercial inexpensive UV-pass filters produced by Subei (model ZWB3) with a thickness of 3mm and a wavelength cut at 420 nm. Its transmission curve was measured and is also shown in figure \ref{fig:transmission}. These filters transmit $45\%$ of Cherenkov light and $\sim 20\%$ direct moonlight.

\begin{figure}[t]
\centering
\includegraphics[width=.35\textwidth]{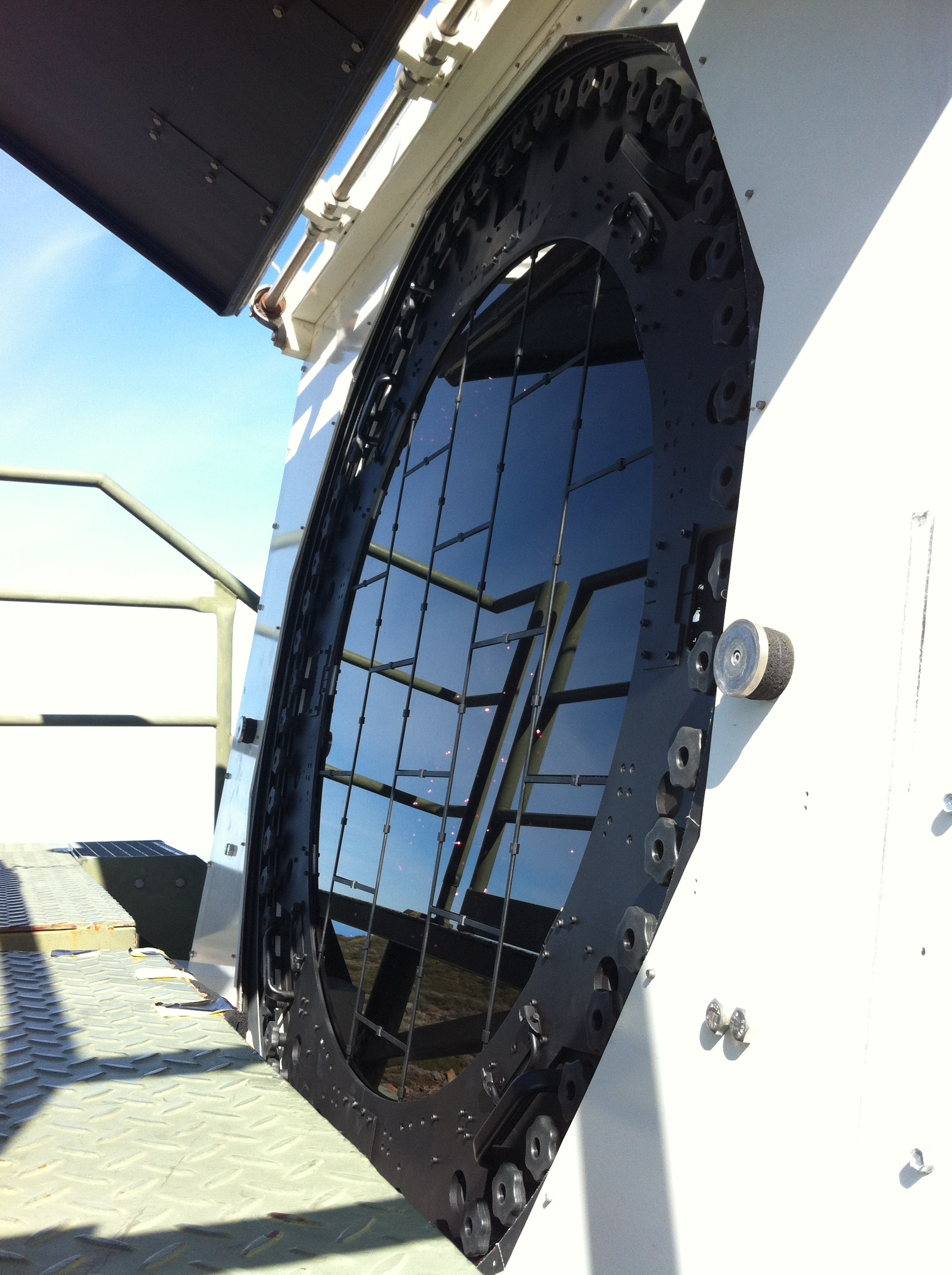}
\includegraphics[width=.6\textwidth]{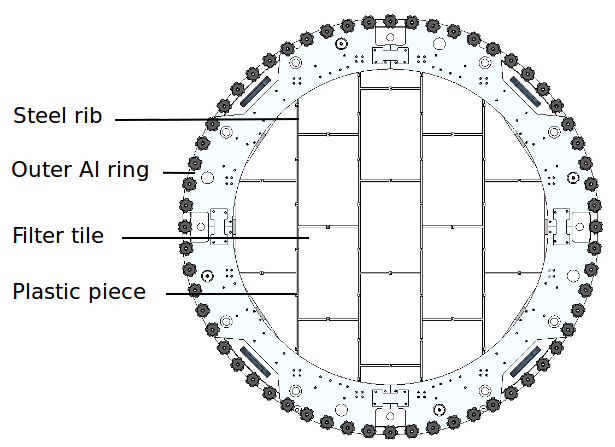}
\caption{On the left, the filters installed in the camera of one of the MAGIC telescopes. On the right, the frame design that holds the filters. The outer Al ring is screwed to the camera. Filter tiles are held by plastic pieces to steel $6 \times 6$ mm$^2$ section ribs.}\label{fig:frame}
\end{figure}

The filters were bought in tiles of $20 \times 30$ cm$^2$, and mounted on a light-weight frame. This frame consists on an outer aluminium ring that is screwed to the PMT camera and steel $6 \times 6$ mm$^2$ section ribs that are placed between the filter tiles (see figure \ref{fig:frame}). The filter tiles are fixed to the ribs by plastic pieces and the space between tiles and ribs is filled with silicone. This gives mechanical stability to the system and prevents light leaks. Two people can mount or dismount the filters in the PMT camera in about 15 minutes.

\section{Data sample and analysis procedure}\label{sec:samples}

After data quality selection a total of almost 15 observation hours of Crab Nebula with the UV-pass filters were recorded. The data were taken in the standard L1-L3 trigger condition\cite{MUP1}, in the so-called wobble mode\cite{Fomin}, with a standard wobble offset of $0.4^\circ$. All the data analysed correspond to zenith angles lower than $35^\circ$.

The data was divided into four samples with different NSB conditions and it is summarized in table \ref{tab1}. The brightness of the sky in each situation is expressed in units of \textit{Dark NSB}. The two samples of highest NSB include situations in which observations without filters are currently impossible in MAGIC.
The mean current measured (DC) in one of the telescopes (MAGIC 1) is also shown in table \ref{tab1} and compared to the expected one under the same brightness conditions, but without filters (Eq. DC). The ratio between both of them depends on how close to the Moon the telescopes are pointing: the background transmission is higher far from the Moon where the diffuse moonlight regime dominates. Close to the Moon the measured DCs with filters can be 5 times lower than the expected one without filters, which is consistent with a moonlight transmission of $20\%$.

To analyse the data and to evaluate the energy threshold, Monte Carlo (MC) simulations are needed. The MC for standard analysis in MAGIC (without filters and in dark conditions) was tuned to include the filters transmission, the shadowing in some pixels that is produced by the ribs of the frame and the increased NSB. With these modified MC simulations, the data has been analysed using the standard MAGIC analysis and reconstructions software, MARS\cite{Aleksic2012b, Zanin}. Due to the relatively high brightness of the sky, the image cleaning settings for each sample were modified with respect to the standard ones\cite{Albert2008a}.

 \begin{table}
\centering
\begin{tabular}{| c | c | c | c | c | c |}
\hline
NSB & $N_{hr}$ & DC in M1 (with filters) & Eq. DC (no filters) & $E_{th}$ & $S_{>300 GeV}$\\

[times \textit{Dark NSB}]  &  [h] &  [$\mu$A]  & [$\mu$A] & [GeV] & [$\%$ C.U.] \\
\hline
   \hline

   \rotatebox[origin=c]{15}{\watermark{PRELIMINARY}}
     2.5 - 5 & 3.4 & 0.5 - 1 & 2 - 3.5 & 175 & $1.03 \pm 0.10$ \\
     5 - 10 & 1.2 & 1 - 2 & 3.5 - 7 & 195 & $1.13 \pm 0.22$  \\
     10 - 20 & 5.7 & 2 - 4 & 7 - 15 & 200 & $1.28 \pm 0.10$ \\
     20 - 40 & 4.1 & 4 - 6 & 15 - 30 & 215 & $1.31 \pm 0.10$ \\
   \hline
 \end{tabular}
 \caption{Samples of Crab Nebula that were taken under different NSB conditions (expressed in terms of the Dark NSB). $N_{hr}$ stands for the total observation time that was achieved under each NSB condition. The third column shows the mean current (DC) measured in MAGIC 1 (M1). The fourth column is the expected measured current in MAGIC 1 under the same NSB conditions, without filters. Last two columns show the preliminary sensitivity and energy threshold for UV-pass filters observations: $E_{th}$ is the energy threshold associated to each NSB range and $S_{>300 GeV}$ its associated sensitivity above 300 GeV, in Crab Units (C.U.). }\label{tab1}
 \end{table}

\section{Performance}\label{sec:performance}

In this section we evaluate the preliminary energy threshold and sensitivity of the telescopes when the filters are installed, for the different NSB conditions that correspond to each of the Crab Nebula samples in table \ref{tab1}.

\subsection{Energy threshold}

To obtain the energy threshold in IACTs one needs to rely on Monte Carlo simulations and to make sure that they describe the data appropriately\cite{MUP}. This is done by comparing distributions of Data and MC and ultimately by checking that the Crab spectrum is well reconstructed using the MC that was specially modified for this analysis. We produced four different sets of MC simulations, each of them adapted to the NSB conditions of each Crab Nebula sample. A common definition of an energy threshold is a peak of an energy distribution plot, obtained from the MC simulations, for a hypothetical source with a spectral index of $-2.6$. Those distribution are built for events that survived the image reconstruction with a typical quality cut of having at least 50 photoelectrons. The energy threshold that is obtained in each range is summarized in table \ref{tab1}. In the darkest situation considered, which corresponds to a sky brightness that goes from 2.5 to 5 times the Dark NSB, the energy threshold is $\sim 175$ GeV, $2.5$ times the achieved one in dark conditions without filters, which is consistent with the fact that the filters cut more than $50\%$ of the Cherenkov light. The energy threshold rises slowly as the NSB increases and it is situated at $\sim 215$ GeV for the brightest situation analysed, which goes up to 40 times the Dark NSB.

\subsection{Sensitivity}

The sensitivity can be defined as the flux of a source giving a $5 \sigma$ significance after 50 h of observation time. The significance is computed using equation 17 from from Li $\&$ Ma with 5 background regions\cite{LiMa}. We apply the conditions $N_{excess} > 10$ and $N_{excess} > 0.05 N_{bgd}$, as explained in \cite{MUP}, where $N_{excess}$ is the number of excess events over a number of background events $N_{bgd}$.

The preliminary integral sensitivity above 300 GeV achieved in each of the NSB ranges considered is summarized in table \ref{tab1}. From the results seems that sensitivity tends to degrade slightly with the NSB. Nevertheless, even in the brightest situation in which the performance was characterized, the sensitivity is close to $1 \% C.U.$, which is $\sim 50\%$ worse than the one that is achieved with MAGIC without filters and in dark conditions.

\begin{figure}[t]
\begin{center}
\subfigure[21:00 UTC]{\includegraphics[scale=0.18]{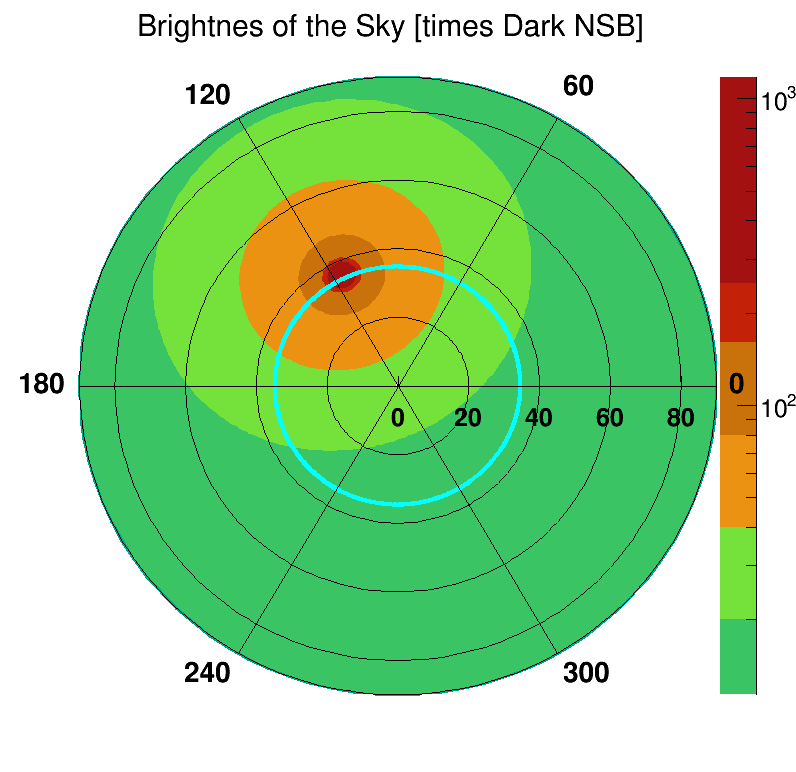}}
\subfigure[00:00 UTC]{\includegraphics[scale=0.18]{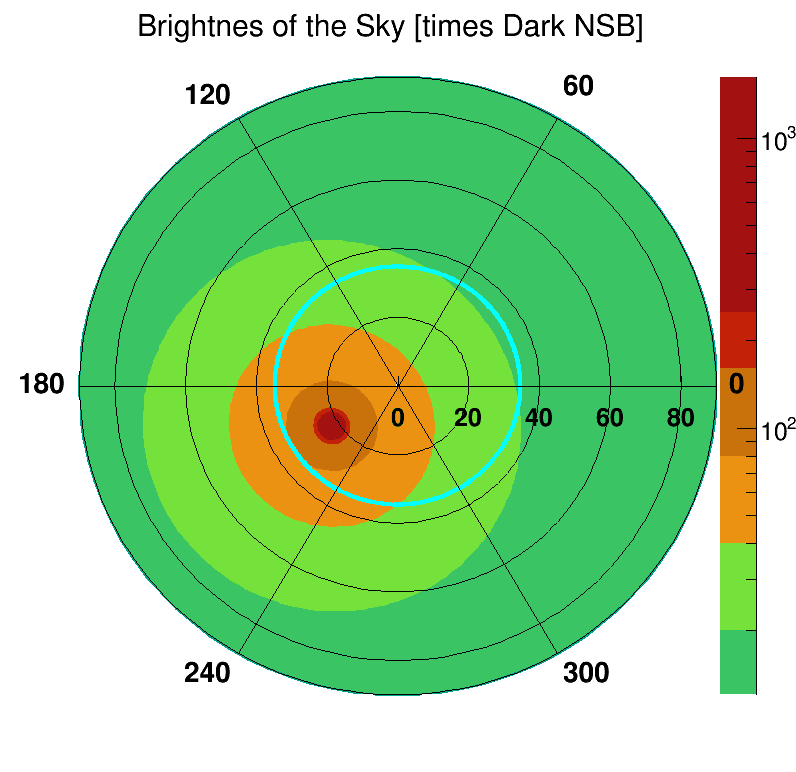}}\\
\subfigure[03:00 UTC]{\includegraphics[scale=0.18]{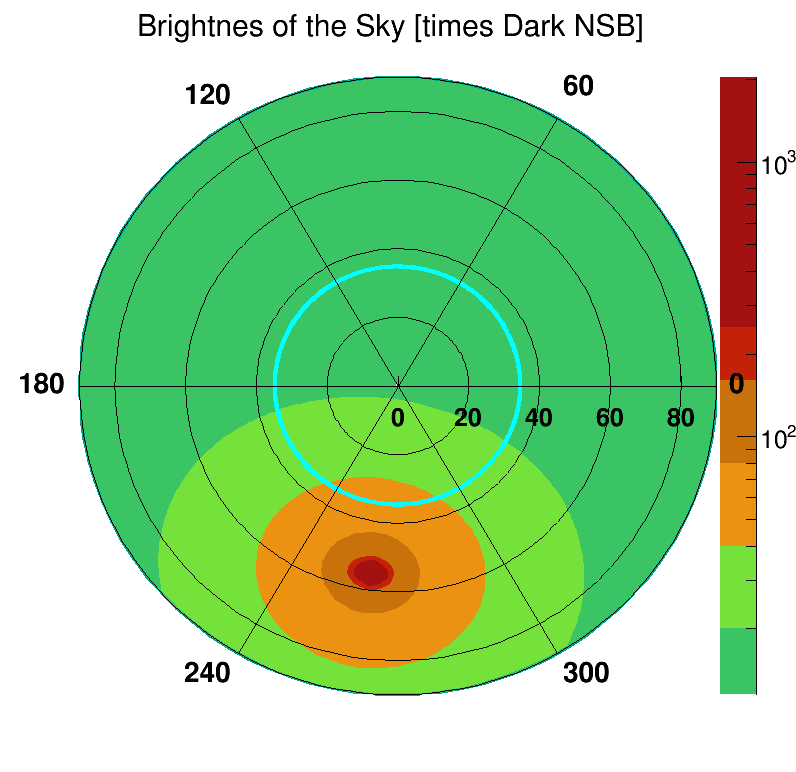}}
\subfigure[05:00 UTC]{\includegraphics[scale=0.18]{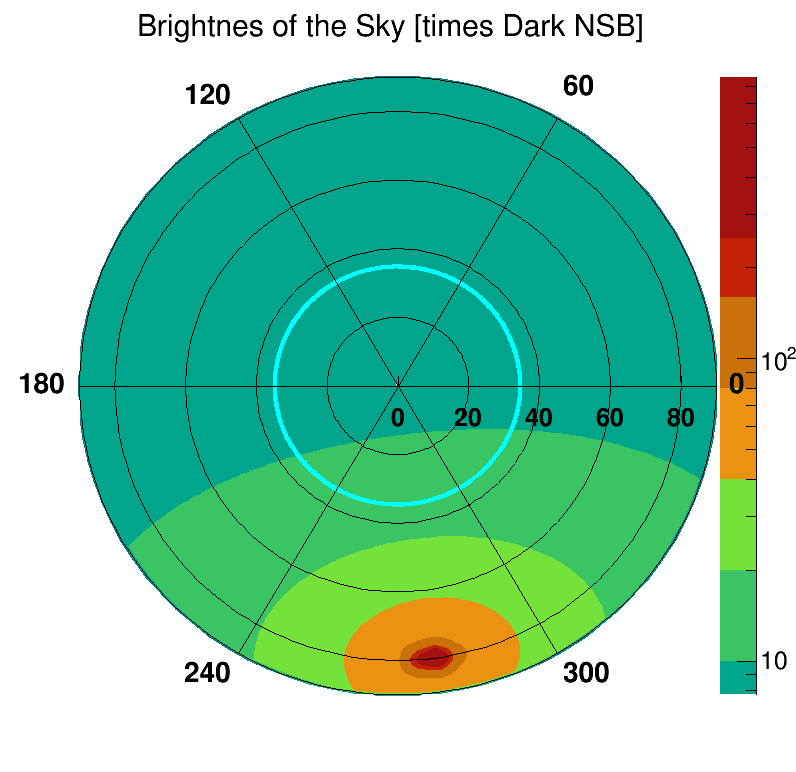}}
\caption{Evolution of the sky brightness during a night of the November full Moon period with a Moon phase of $\sim 95\%$. The color scale gives the brightness in units of times the Dark NSB. Red and orange zones correspond to regions in which the performance is still unknown. Green zones, to those in which the performance was characterized.}\label{fig:SkyBr}
\end{center}
\end{figure}

\section{Discussion}\label{sec:conclusions}

With the use of UV-pass filters observations with MAGIC are feasible in situations in which it was impossible before. Observations can be performed in conditions that are safe for the PMTs up to a distance of $5^\circ$ away from the Moon during full Moon, which means that the telescopes can point almost at any position in the sky, whichever the Moon phase is. This means that with the filters installed, the telescopes can now be operated during the full Moon period.

However, the sensitivity is $\sim 50 \%$ worse than the standard one for MAGIC without filters. On the other hand, the performance is characterized up to 40 times the Dark NSB (which still corresponds to conditions in which observations without filters are impossible and to situations in which many of the target sources are found) but it is unknown beyond that limit, essentially because of lack of Crab Nebula data in those conditions. In figure \ref{fig:SkyBr} we can see, as an example, what would be the brightness of the sky\cite{Britzger} during a night of the November 2015 full Moon period in which the Moon phase is $95\%$. Red and orange regions correspond to brightness in which the performance is yet not characterized. Different tones of green correspond to the different brightness ranges in which the performance is known. While the Moon is low in the sky, almost the whole sky is accessible. When the Moon is close to zenith the portion of the sky of unknown performance is larger.

The main result is that with the use of the UV-pass filters, MAGIC duty cycle could be increased by $30 \%$ by taking data over the entire nights of the full Moon period and the two nights before and after it. This makes the filters useful for discovering new sources and for monitoring of variable sources, as well as for deep observations of hard sources.

An interesting project that could take profit from the increased duty cycle and from the possibility of observing close to the Moon is to observe the deficit in the cosmic ray flux produced by the presence of the Moon (the so-called ``Moon shadow'', see \cite{Colin}). About 100 observation hours per year  with filters could be used for this project. However, as it implies pointing very close to the Moon, the sky brightness is much higher than in the situations in which the performance is known (it can be as high as 200 times the Dark NSB), which makes it much more challenging.

\section{Acknowledgments}

We would like to thank
the Instituto de Astrof\'{\i}sica de Canarias
for the excellent working conditions
at the Observatorio del Roque de los Muchachos in La Palma.
The financial support of the German BMBF and MPG,
the Italian INFN and INAF,
the Swiss National Fund SNF,
the ERDF under the Spanish MINECO (FPA2012-39502), and
the Japanese JSPS and MEXT
is gratefully acknowledged.
This work was also supported
by the Centro de Excelencia Severo Ochoa SEV-2012-0234, CPAN CSD2007-00042, and MultiDark CSD2009-00064 projects of the Spanish Consolider-Ingenio 2010 programme,
by grant 268740 of the Academy of Finland,
by the Croatian Science Foundation (HrZZ) Project 09/176 and the University of Rijeka Project 13.12.1.3.02,
by the DFG Collaborative Research Centers SFB823/C4 and SFB876/C3,
and by the Polish MNiSzW grant 745/N-HESS-MAGIC/2010/0.

\end{document}